# A Low-latency Secure Data Outsourcing Scheme for Cloud-WSN

Jing Li[1], Zhitao Guan[1], Xiaojiang Du[2], Zijian Zhang[3], Zhenyu Zhou[1]
1. School of Control and Computer Engineering, North China Electric Power University, China, lijing_immanuel@163.com, guan@ncepu.edu.cn, zhenyu_zhou@ncepu.edu.cn
2. Department of Computer and Information Science, Temple University, Philadelphia PA, USA, dxj@ieee.org
3. School of Computer, Beijing Institute of Technology, Beijing, China, zhangzijian@bit.edu.cn

*Abstract*—**With the support of cloud computing, large quantities of data collected from various WSN applications can be managed efficiently. However, maintaining data security and efficiency of data processing in cloud-WSN (C-WSN) are important and challenging issues. In this paper, we present an efficient data outsourcing scheme based on CP-ABE, which can not only guarantee secure data access, but also reduce overall data processing time. In our proposed scheme, a large file is divided into several data blocks by data owner (DO) firstly. Then, the data blocks are encrypted and transferred to the cloud server in parallel. For data receiver (DR), data decryption and data transmission is also processed in parallel. In addition, data integrity can be checked by DR without any master key components. The security analysis shows that the proposed scheme can meet the security requirement of C-WSN. By performance evaluation, it shows that our scheme can dramatically improve data processing efficiency compared to the traditional CP-ABE method.**

## I. INTRODUCTION

In order to enhance the scalability of the WSN, some studies focus on combining cloud computing and wireless sensor networks [1]. Security in WSN has been studied in a lot of literature, e.g., [2-5].With the support of cloud computing, cloud-WSN (C-WSN) can be constructed. It can be viewed as a special kind of Heterogeneous Sensor Networks [6-8]. C-WSN has been used in many applications, such as smart grid. In smart grid, there are lots of WSN based applications, including power transmission line monitoring, smart metering & smart home, power consumption information gathering, etc. Mass data from different WSN applications are collected and stored in cloud servers. Then, different type of data receivers (DR) will access the data according to their own access rights.

However, there are still several problems and challenges in C-WSN. First, data processing efficiency should be considered due to large amount of data is encrypted/decrypted and transferred in C-WSN. Second, data security and data privacy must be kept in mind. To solve these problems, we propose an efficient data outsourcing scheme, which can dramatically enhance data processing efficiency in C-WSN without loss of data security and data privacy. Main contributions of this paper can be summarized as follows: 1) We propose a block-encryption method which allows a large file to be encrypted/decrypted and transmitted in blocks in parallel. 2) The data receiver can check data integrity without any master key components. 3) We give the security analysis and performance evaluation, which prove that security and performance of our scheme are no weaker than that of traditional scheme.

The rest of this paper is organized as follows. Section 2 introduces the related work. In section 3, some preliminaries are given. In section 4, our scheme is stated. In section 5, security analysis is given. In Section 6, the performance of our scheme is evaluated. In Section 7, the paper is concluded.

## II. RELATED WORK

In [9], the ciphertext's encryption policy is associated with a set of attributes, and the data owner can be offline after data is encrypted. In [10] and [11], Key Policy Attribute-Based Encryption (KP-ABE) scheme and Ciphertext Policy Attribute-Based Encryption (CP-ABE) scheme were proposed respectively. In KP-ABE, the encryption policy is also associated with a set of attributes, which are organized into a tree structure (named access tree) by users. In CP-ABE, the data owner constructs the access tree using visitors' identity information.

In [12], Yu et al tried to achieve secure, scalable, and fine-grained access control in a cloud environment. Wang et al proposed an access control scheme based on CP-ABE, which was also secure and efficient in a cloud environment [13]. In [14], Yadav and Dave presented an access model based on CP-ABE which could provide a remote integrity check by way of augmenting secure data storage operations. In addition, there are still some researches and applications of access control in practical problems, e.g., [15, 16]. Hei et al applied the access control scheme to the medical field, in order to solve the various problems in practice. They firstly discussed and studied how to detect the two attacks against insulin pump systems via wireless links in [15], and the feasibility of the scheme is proved by experiments. Then in [16], Hei et al considered the Implantable Medical Devices (IMD) security, and proposed a light-weight secure access control scheme. With the small computation overhead, the scheme can be applied to solve the security problems in emergency situations.

The similarity between the existing works and ours is that we are all based on ABE method. While, we make an innovation and improvements based on CP-ABE. In order to enhance the data processing efficiency in C-WSN, we propose a novel partition method based on CP-ABE to enable the data encryption and data transmission to be processed in parallel.

## III. PRELIMINARIES

### A. Bilinear Maps and Complexity Assumptions

Let $G_0$ and $G_1$ be two multiplicative cyclic groups of prime order $p$ and $g$ be the generator of $G_0$. The bilinear map $e$ is, $e: G_0 \times G_0 \to G_1$, for all $a, b \in \mathbb{Z}_p$:

- Bilinearity: $\forall u, v \in G_1, e(u^a, v^b) = e(u, v)^{ab}$
- Non-degeneracy: $e(g, g) \neq 1$
- Symmetric: $e(g^a, g^b) = e(g, g)^{ab} = e(g^b, g^a)$

**Definition** Discrete Logarithm (DL) Problem:

Let $G$ be a multiplicative cyclic group of prime order $p$ and $g$ be its generator, given $y \in_R G$ as input, try to get $x \in \mathbb{Z}_p$ that $y = g^x$.

The DL assumption holds in $G$ if it is computationally infeasible to solve DL problem in $G$.

### B. Structure in Ciphertext-policy Attribute Based Encryption (CP-ABE)

To achieve fine-grained access control, we utilize the Ciphertext Policy Attribute-Based Encryption scheme (CP-ABE) [4], which the access structure is illustrated by an access tree. Leaves of the tree are associated with descriptive attributes, and each interior node is a relation function, such as AND ($n$ of $n$), OR (1 of $n$), and n of m ($m>n$).

Let $\mathcal{T}$ be an access tree, and the root node is denoted by $\mathcal{R}$. At the beginning of the encryption, we will conduct a polynomial for each node from top to bottom, while the decryption order is reverse.

To retrieve the secret, we define the Lagrange coefficient $\Delta_{i,S}$ as follows:

For $i \in \mathbb{Z}_p$, and for $\forall x \in S$,

$$\Delta_{i,S(x)} = \prod_{j \in S, j \neq i} \frac{x-j}{i-j}.$$

## IV. DESCRIPTION OF OUR SYSTEM

### A. System Model

In our system, both Data Owners (denoted as DO) and Data Requester/Receivers (denoted as DR) are users, as shown in Fig.1. The Trusted Authority (TA) is a trusted party to generate Public Keys (PK), Master Keys (MK) for data encryption and Secret Key (SK) for each DR. Cloud Servers(CS) are assumed to be semi-trusted. We employ them to be in charge of storing our encrypted data.

### B. Encryption/Decryption and transmission in parallel

We partition the file and the access tree into several blocks, which allows each data block to be encrypted and decrypted independently, paralleling to transmission and computation of the different data blocks. This reduces the response time of severs and shortens the DR's wait time.

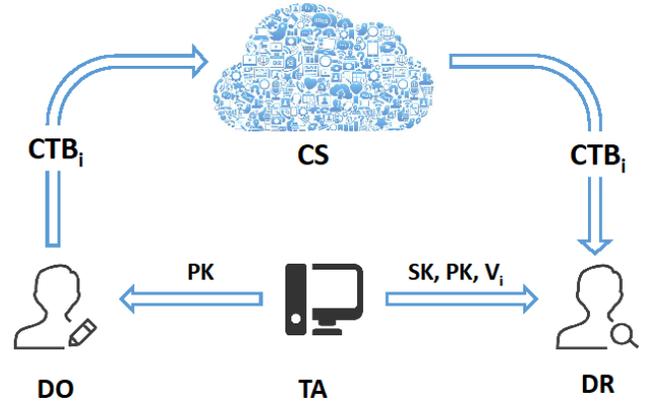

Fig. 1. System Model

As long as one block has been encrypted, it can be transmitted, which allows another block to be encrypted at the same time. As shown in Fig. 2, we make $T_1$ denote the total time of encryption and transmission of our scheme and $T_2$ denote the time of CP-ABE. The diagram below shows the advantage of the improved scheme.

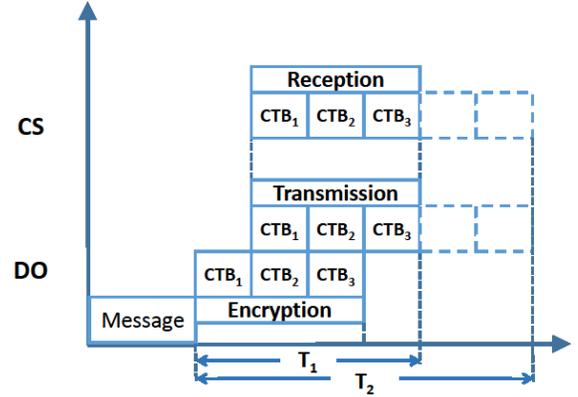

Fig. 2. The encryption-transmission time comparison between our scheme and CP-ABE

As shown in Fig. 3, the similar situation occurs in decryption. DR decrypts one block, while others are being transmitting over the internet. We make $T_1$ denote the total time of transmission and decryption of our scheme and $T_2$ denote the time of CP-ABE. The diagram shows the advantage of the improved scheme.

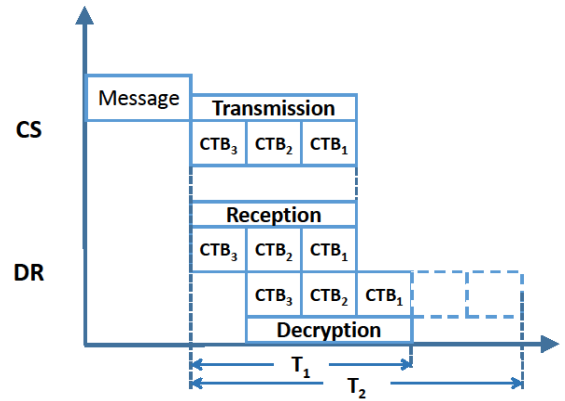

Fig. 3. The transmission-decryption time comparison between our scheme and CP-ABE

## C. Algorithms

### 1) Setup

The TA will run this setup algorithm and generate a set of public parameters. First, it chooses a bilinear group $G_0$ of prime order $p$ with generator $g$ and several random exponents: $\alpha, \beta, q \in \mathbb{Z}_p$. We introduce hash functions $H_v(), H_{att}()$ for plaintext and all of the attributes. The public key and the master key are published as:

$$PK = \{G_0, g, h = g^\beta, e(g,g)^\alpha\}$$
$$MK = \{\beta, g^\alpha, q, k\} \quad (1)$$

### 2) Key Generation

DR should legally register to the TA, which will evaluate his attributes and generate a corresponding SK for him. The algorithm is as follows:

**Key_Gen**(*PK, MK, S*) →*SK*

The set of attributes *S* will be the input and a corresponding key is the output. This algorithm first chooses $r \in \mathbb{Z}_p$ and $r_j \in \mathbb{Z}_p$ at random for each attribute $j \in S$. $q$ is one of the MK components. Then it computes the key as:

$$SK = \begin{pmatrix} D = g^{\frac{\alpha+r}{\beta}}, \hat{D} = g^{rq}, \\ \forall j \in S, D_j = g^r H(j)^{r_j}, D_j' = g^{r_j} \end{pmatrix} \quad (2)$$

### 3) Encryption

This process involves several specific steps. DO first computes the message *M* with a hash function $H_v()$ and MK, then he will partition the message *M* into several data blocks (DB) according to the level number of its access tree. Next, he will perform a *XOR* operation of two different data blocks. Finally, these data blocks will be encrypted and transmitted one by one.

**Data_Verification**(*M, PK, MK*)→ $\check{C}$

This algorithm gets a message *M* and a hash function as inputs, computes and outputs:

$$\check{C} = H_v(M)^k \quad (3)$$

It will be one components of the first ciphertext block ($CTB_1$), and DR utilizes this parameter to verify the correctness of his decryption results.

**Data_Partition**(*M, n*)→*DB$_i$*

DO constructs the access tree and gets its level number *n*. This algorithm takes the message and the number as input, partitioning and computing as follows:

$$M \to M_1, M_2, M_3 ... M_n \quad (4)$$

$$DB_1 = M_1, DB_2 = M_1 \oplus M_2, DB_3 = M_2 \oplus M_3, ...,$$
$$DB_n = M_{n-1} \oplus M \quad (5)$$

Finally, it outputs $DB_1, DB_2, DB_3, ... DB_n$.

**DB_Encryption**(*DB$_i$, AS$_i$, PK*)→*CTB$_i$*

This algorithm first selects $s_1 \in \mathbb{Z}_p$ for $DB_1$ and $s_2 \in \mathbb{Z}_p$ for $DB_2$ and generate the polynomials for the first level (it contains the root node $\mathcal{R}$) of current access tree, which is similar to that of [4]. It will compute as follows:

$$Sec_2 = g^{\frac{s_2}{q}} \quad (6)$$

$$\tilde{C}_1 = (DB_1 \| Sec_2) e(g,g)^{\alpha s_1} \quad (7)$$

$$C_1 = h^{s_1} \quad (8)$$

We make each *CTB* contain the secret of the lower level to ensure that the corresponding *CTB* can be successfully decrypted.

If the first level of the access tree contains leaf nodes, let $Y_{l-1}$ be the set of leaf nodes, computing as follows:

$$\forall y \in Y_{l-1}, \hat{C}_{1,y} = g^{q_{1,y}(0)}, \hat{C}_{1,y}' = H_a(att(y))^{q_{1,y}(0)} \quad (9)$$

All the polynomials ($q_{n1}(x), q_{n2}(x)...$) that related to the interior nodes of the current level will be classified as the next level of the access structure $AS_{i+1}$.

The complete form of the first ciphertext block ($CTB_1$) is as follows:

$$CTB_1 = (\tilde{T}_{l-1}, \tilde{C}_1, C_1, \check{C}, \forall y \in Y_{l-1}, \hat{C}_{1,y}, \hat{C}_{1,y}') \quad (10)$$

Then the current *CTB* will be transmitted over the network to the CS.

Except for the $DB_1$, the calculation process of other DBs is the same. We choose $DB_2$ as an example of the encryption. First, the algorithm gets $s_2$ from the previous step and selects $s_3 \in \mathbb{Z}_p$ for the next level and computes as follows:

$$Sec_3 = g^{\frac{s_3}{q}} \quad (11)$$

$$\tilde{C}_2 = (DB_2 \| g^{\frac{s_3}{q}}) e(g,g)^{\alpha s_2} \quad (12)$$

$$C_2 = h^{s_2} \quad (13)$$

This level may contain several interior nodes. According to these polynomials, for each interior node $N_j$, the algorithm computes:

$$\Delta s_j = s_2 - q_{n_j}(0) \quad (14)$$

$$\Delta C_{2,j} = g^{\frac{\Delta s_j}{q}} \quad (15)$$

Similarly, the algorithm will generate a polynomial for each interior child node, which consists of the next level of *AS*. If the current level contains leaf nodes, let $Y_{l-2}$ be the set of leaf nodes, computing as follows:

$$\forall y \in Y_2, \hat{C}_{2,y} = g^{q_{2,y}(0)}, \hat{C}_{2,y}' = H_a(att(y))^{q_{2,y}(0)} \quad (16)$$

The complete form of the current ciphertext block (*CTB*) is as follows:

$$CTB_2 = \left(\tilde{T}_{l-2}, \tilde{C}_2, C_2, \Delta C_{2,j}, \forall y \in Y_{l-2}, \hat{C}_{2,y}, \hat{C}_{2,y}{}'\right) \quad (17)$$

The construction of the other *CTBs* is also the same.

$$CTB_i = \left(\tilde{T}_{l-i}, \tilde{C}_i, C_i, \Delta C_{i,j}, \forall y \in Y_{l-i}, \hat{C}_{i,y}, \hat{C}_{i,y}{}'\right) \quad (18)$$

*4) Decryption*

Each *CTB* is encrypted under only one level rather than the whole access tree, here will not be any recursion algorithm within the calculation. Our scheme allows DRs to wait the other *CTBs* while decrypting one *CTB*.

**Decrypt_LeafNode** ($CTB_i$, $SK$, $z$)→$F_z$ or $\perp$

We define the algorithm that takes *CTB* and *SK* as inputs. As long as a DR gets one *CTB* from CS, he will perform this algorithm:

If the node $z$ is a leaf node, let $j=att(z)$, if $j \in S$, then,

$$\begin{aligned} F_z &= \frac{e(D_j, \hat{C}_{i,y})}{e(D_j{}', C_{i,y}{}')} \\ &= \frac{e(g^r H_a(j)^{rj}, g^{q_z(0)})}{e(g^{rj}, H_a(att(z))^{q_z(0)})} \\ &= e(g,g)^{rq_z(0)} \end{aligned} \quad (19)$$

Else, it returns $\perp$.

All of the leaf nodes contained in this $CTB_i$ will be calculated, and the outputs will be stored.

**Decrypt_InteriorNode** ($F_z$, $S_x$)→$F_x$ or $\perp$

Let $x$ be the parent node of these nodes $z$, let $S_x$ be the set of its child nodes. If $\forall z \in S_x, F_z = \perp$, the function returns $\perp$. Else, we can get the parent node value by computing:

$$\begin{aligned} F_x &= \prod_{z \in S_x} F_z^{\Delta_{i,S_x'}(0)}, \text{ where } \begin{cases} n = index(z) \\ S_x' = \{index(z) : z \in S_x\} \end{cases} \\ &= \prod_{z \in S_x} (e(g,g)^{r \cdot q_z(0)})^{\Delta_{n,S_x'}(0)} \\ &= \prod_{z \in S_x} (e(g,g)^{r \cdot q_{parent(z)}(index(z))})^{\Delta_{i,S_x'}(0)} \\ &= \prod_{z \in S_x} (e(g,g)^{r \cdot q_x(n)})^{\Delta_{i,S_x'}(0)} \\ &= e(g,g)^{r \cdot q_x(0)} \end{aligned} \quad (20)$$

**Decrypt_DB**($CTB_i$, $SK$, $F_x$)→$DB_i$, $Sec_{i+1}$

DR will repeatedly call those two functions to gets all of the parent nodes value. If $i \neq 1$, as long as one of them is retrieved (assume that the parent node $j$ is retrieved), this algorithm computes as follows:

Set $A$ as the intermediate result.

$$\begin{aligned} A &= F_x e(\Delta C_{i,j}, \hat{D}) = e(g,g)^{rq_x(0)} e(g^{\frac{\Delta s_j}{q}}, g^{rq}) \\ &= e(g,g)^{rq_x(0)} e(g,g)^{r\Delta s_j} \\ &= e(g,g)^{rs_i} \end{aligned} \quad (21)$$

$$\frac{\tilde{C}_i A}{e(C_1, D)} = \frac{(DB_i \| g^{\frac{s_{i+1}}{q}}) e(g,g)^{(\alpha+r)s_i}}{e(h^{s_i}, g^{\frac{\alpha+r}{\beta}})} = (M_i \oplus M_{i-1}) \| g^{\frac{s_{i+1}}{q}} \quad (22)$$

$$DB_i = M_i \oplus M_{i-1} \quad (23)$$

$$Sec_i = g^{\frac{s_{i+1}}{q}} \quad (24)$$

If $i=1$, the algorithm computes as follows:

$$\frac{\tilde{C}_1 F_R}{e(C_1, D)} = \frac{(DB_1 \| g^{\frac{s_2}{q}}) e(g,g)^{(\alpha+r)s_1}}{e(h^{s_1}, g^{\frac{\alpha+r}{\beta}})} = M_1 \| g^{\frac{s_2}{q}} \quad (25)$$

$$DB_1 = M_1 \quad (26)$$

$$Sec_2 = g^{\frac{s_2}{q}} \quad (27)$$

Obviously, there is no guarantee that all *CTBs* will be successfully decrypted. Thus, all these *CTBs*, *DBs* and *Sec* parameters will be stored as inputs of the algorithm Decrypt_M($S_{CTB}$, $S_{DB}$, $S_{Sec}$, $SK$).

**Decrypt_M**($S_{CTB}$, $S_{DB}$, $S_{Sec}$, $SK$)→$M$ or $\perp$

If $DB_1 \notin S_{DB}$, which means his attributes don't satisfy the access policy. Even with some other decrypted *DBs*, M cannot be retrieved at all. This algorithm outputs $\perp$.

If $S_{CTB} \neq \varnothing$, for $\forall CTB_i \in S_{CTB}$, this algorithm computes as follows:

$$\begin{aligned} \frac{\tilde{C}_i e(Sec_i, \hat{D})}{e(C_1, D)} &= \frac{(DB_i \| g^{\frac{s_{i+1}}{q}}) e(g,g)^{\alpha s_i} e(g^{\frac{s_i}{q}}, g^{rq})}{e(h^{s_i}, g^{\frac{\alpha+r}{\beta}})} \\ &= M_i \| g^{\frac{s_{i+1}}{q}} \end{aligned} \quad (28)$$

Else, for $\forall DB_j \in S_{DB}$, the message M can be retrieved as follows:

$$\begin{aligned} M_1 &= M_1, M_2 = DB_2 \oplus M_1, M_3 = DB_3 \oplus M_2, \ldots \\ M_n &= DB_n \oplus M_{n-1} \end{aligned} \quad (29)$$

$$M = M_1 \| M_2 \| M_3 \ldots \| M_n \quad (30)$$

Finally, it outputs *M*.

*5) Verification*

In our scheme, a complete message *M* is partitioned and encrypted into several *CTBs*. We define an algorithm to allow DR to verify the correctness of decryption. Before that, TA will get $\tilde{C} = H_v(M)^k$ from CS and randomly select $t \in \mathbb{Z}_p$, computing as follows:

$$H_v(M)^k \rightarrow \left(H_v(M)^k\right)^{\frac{t}{k}} \qquad g \rightarrow g^t \quad (31)$$

These two parameters will be stored as a tuple $V:\{V_1, V_2\}$ and transmitted to DR.

**Verify_M**(*M*, *V*, *PK*)→*True* or *False*

First, DR will utilize the Hash function to calculate this $M$:

$$M \to H_v(M) \tag{32}$$

Verify that:

$$e(H_v(M), V_2) = e(V_1, g) \tag{33}$$

If (33) exits, this algorithm will output *True*. Else, it will output *False*.

## V. SECURITY ANALYSIS

### A. System Security

We parallel the transmission and computation by partitioning the data and access tree into chunks, which reduces the response time of severs and the DR's waiting time. The ways to encrypt and decrypt affect the security of the system.

Theorem: The security of our system is no weaker than that of [4].

Proof: Although we partition the access tree and the message, the ways used for encryption and decryption is analogous to that in [4]. In our system, $DB(M_{i-1} \oplus M_i)$ is multiplied by $e(g,g)^{\alpha s}$.

In order to decrypt, an attacker clearly retrieves $s$ either by decrypting the corresponding level of access tree or obtaining its $Sec_i$. Even it gets all the *DBs* by the latter way, it cannot retrieve any $M_i$ without $M_1$. However, the only way to get $M_1$ is to compute the access tree.

### B. Partition

A complete tree will be partitioned into several levels, and each data block (DB) will be masked by a unique random number $s_i$ chosen for one of these levels, which is different from any of the interior nodes value. We lower the difficulty of decryption, namely, as long as one interior node value contained in this level is retrieved, the DB can be successfully decrypted. It means that even the DR's attributes don't satisfy the access policy, he is able to get access to some of the DBs.

Theorem: Partition doesn't impact the security of our scheme.

Proof: Compared with a complete tree, each level contains partial access policy. There is no other information being inserted into these partitioned ones.

As for DBs, all of them are preprocessed except for the first one. Without the first one, even all the others are acquired, none of them can be retrieved. The only way to access to the first DB is to satisfy its access policy, which is the same as that of [4]. Thus, to partition the message and the tree doesn't impact the security of the proposed scheme.

## VI. PERFORMANCE EVALUATION

### A. Numerical analysis

In this paper, we partition a large file into several data blocks, which can be encrypted, decrypted, and transmitted concurrently. Therefore, we can be able to administer the encryption/decryption and the transmission of different blocks in parallel. Now, we give the efficiency analysis according to the process shown in Fig. 4. The time cost in each step is shown in Table I.

TABLE I. THE PARAMETERS IN PERFORMANCE EVALUATION

| Symbols | Description |
|---|---|
| $n$ | M is partitioned into n blocks. |
| $ET_M$ | Encryption time of a whole file |
| $ET_i$ | Encryption time of the a data block i. |
| $TT_M$ | Transmission Encryption time of a whole file |
| $TT_i$ | Transmission time of the a data block i. |
| $DT_M$ | Decryption time of the whole file. |
| $DT_i$ | Decryption time of the a data block i. |

*1) Encryption-Transmission*

The total time of traditional scheme [4] is,

$$ET_M + TT_M \tag{34}$$

If $TT_i > ET_i$, the total time of our scheme is,

$$ET_1 + \sum TT_i \approx ET_1 + TT_M \tag{35}$$

Else, the total time is,

$$TT_n + \sum ET_i \approx ET_M + TT_n \tag{36}$$

We set $\Delta T$ denotes the difference:

$$\Delta T \approx ET_M + TT_M - ET_1 - TT_M \approx ET_M - ET_1$$
$$\Delta T \approx ET_M + TT_M - ET_M - TT_n \approx TT_M - TT_n \tag{37}$$

For all of the data we suppose that there is no delay in the process of block encryption/decryption or transmission. Therefore, there is no time gap between two consecutive blocks in data block encryption, decryption, and transmission. Thus, in either case, $\Delta T > 0$.

*2) Transmission-Decryption*

The total time of traditional scheme [4] is,

$$TT_M + DT_M \tag{38}$$

If $TT_i > ET_i$, the total time of our scheme is,

$$\sum TT_i + DT_n \approx TT_M + DT_n \tag{39}$$

Else, the total time is,

$$TT_1 + \sum DT_i \approx DT_M + TT_1 \tag{40}$$

We set $\Delta T$ denotes the difference :

$$\Delta T \approx DT_M + TT_M - DT_n - TT_M \approx DT_M - DT_n$$
$$\Delta T \approx DT_M + TT_M - DT_M - TT_1 \approx TT_M - TT_1 \tag{41}$$

Ideally, $DT_M > DT_n$, $TT_M > TT_1$.

Thus, in either case, $\Delta T > 0$.

### B. Experimental results

To evaluate the performance of our system, we implemented a testing environment with the help of the cpabe-toolkit [17].

Encryption and transmission can be concurrent based on data blocks. Compared to [4], the total time of encryption and transmission can be reduced. The results shown in Fig. 4(a) indicate that the total time drops as the size of message increases.

We encrypt the message under an access tree that contained ten levels and a hundred leaf nodes.

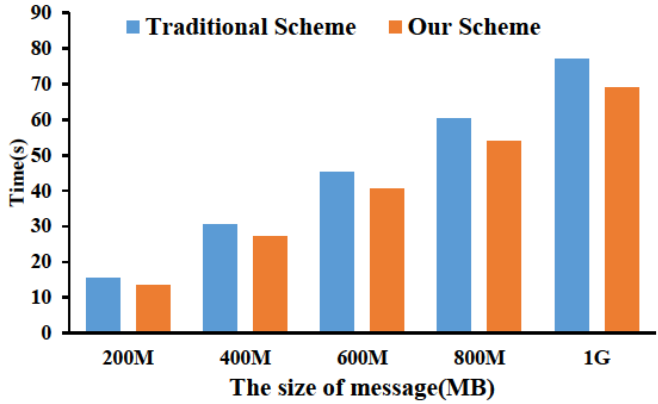

Fig. 4(a). Comparison of the encryption-transmission time between the tradition scheme in [4] and ours, when the size of message grows.

To a certain extent, different decryption keys can affect the decryption time overhead. We use the same key to conduct the experiment. Compared with the transmission time, the decryption time is relatively short. So as shown in Fig. 4(b), we select the difference to indicate the comparison between the traditional scheme and ours.

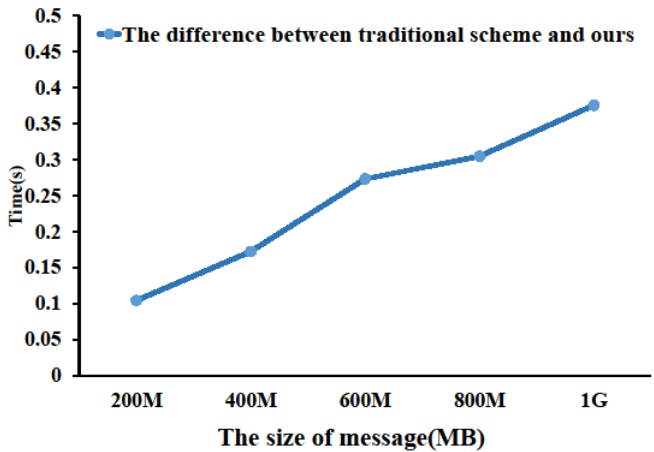

Fig. 4(b). The transmission-decryption time difference between the tradition scheme in [4] and ours, when the size of message grows.

## VII. CONCLUSION

It is a very hot topic to provide security for C-WSN. C-WSN is suitable for being applied in industrial system, such as smart grid. In C-WSN based applications, mass data is collected by WSN and managed by the cloud efficiently. However, maintaining data security and system efficiency meanwhile in C-WSN is a challenging issue. Existing related schemes seldom take efficiency of data processing into consideration. To solve this problem, this paper proposes an efficient data outsourcing scheme based on CP-ABE, which can not only guarantee secure data access, but also reduce overall data processing time. The security analysis shows that the proposed scheme can meet the security requirement of C-WSN. The evaluations show the advantages on the efficiency of data processing, especially for data encryption and data transmission.

ACKNOWLEDGMENT

This work is partially supported by Natural Science Foundation of China under grant 61402171, the Fundamental Research Funds for the Central Universities under grant 2016MS29, as well as by the US National Science Foundation under grant CNS-1564128 and the Qatar National Research Fund under grant NPRP 8-408-2-172.